\documentclass{ws-ijmpd}
\usepackage[super,compress]{cite}
\usepackage{mathrsfs}

\begin{document}


%
\catchline{}{}{}{}{}
%

\title{COMPACT OBJECTS IN GENERAL RELATIVITY: FROM BUCHDAHL
STARS TO QUASIBLACK HOLES}

\author{JOS\'E P. S. LEMOS}

\address{Centro de Astrof\'isica e Gravita\c c\~ao - CENTRA,
Departamento de F\'{\i}sica, 
Instituto Superior T\'ecnico - IST, Universidade  de Lisboa - UL,
Avenida~Rovisco Pais 1, 1049-001 Lisboa, Portugal\\
joselemos@ist.utl.pt}

\author{OLEG B. ZASLAVSKII}

\address{Astronomical Institute of Kharkov V.~N. 
Karazin National University, 35 Sumskaya St.,
Kharkov, 61022, Ukraine\\
ozaslav@kharkov.ua}

\maketitle


\begin{abstract}
A Buchdahl star is a highly compact star for which the boundary radius
$R$ obeys $R=\frac98 r_+$, where $r_+$ is the gravitational radius of
the star itself. A quasiblack hole is a maximum compact star, or more
generically a maximum compact object, for which the boundary radius $R$
obeys $R=r_+$. Quasiblack holes are objects on the verge of becoming
black holes. Continued gravitational collapse ends in black holes and
has to be handled with the Oppenheimer-Snyder formalism. Quasistatic
contraction ends in a quasiblack hole and should be treated with
appropriate techniques. Quasiblack holes, not black holes, are the
real descendants of Mitchell and Laplace dark stars. Quasiblack holes
have many interesting properties. We develop the concept of a
quasiblack hole, give several examples of such an object, define what
it is, draw its Carter-Penrose diagram, study its pressure properties,
obtain its mass formula, derive the entropy of a nonextremal
quasiblack hole, and through an extremal quasiblack hole give a
solution to the puzzling entropy of extremal black holes.
\end{abstract}




\section{Introduction}

In general relativity, a compact object is a body whose radius $R$ is
not much larger than its own gravitational radius $r_+$.  Compact
objects are realized in compact stars.  The concept of a compact
object within general relativity achieved full form with the work of
Buchdahl\cite{buch} where it was proved on quite general premises that
for any nonsingular static and spherically symmetric perfect fluid
body configuration of radius $R$ with a Schwarzschild exterior, the
radius $R$ of the configuration is bounded by $R\geq\frac98\,
r_+$, with $r_+=2m$ in this case, $m$ being the
spacetime mass, and we use units in which the constant
of gravitation and the velocity of light are set
equal to one.  Objects with $R=\frac98\,r_+$ are called
Buchdahl stars, and are highly compact stars.  A Schwarzschild star,
i.e., what is called the Schwarzschild interior
solution\cite{schwarz2}, with energy density $\rho$ equal to a
constant, is a realization of this bound.  Schwarzschild stars can
have any
relatively large
radius $R$ compared to their gravitational radius $r_+$, but
when the star has radius $R=\frac98\,r_+$, i.e., it is a
Buchdahl star, the inner pressure goes to infinity and the solution
becomes singular at the center, solutions with smaller
radii $R$ being even more singular.
From here, one can infer that when the star becomes a Buchdahl star,
i.e., its radius $R$, by a quasistatic process say, achieves
$R=\frac98\,r_+$, it surely collapses.
A neutron star, of
radius of the order $R=3r_+$, although above the
Buchdahl limit, is certainly a compact star, and its apparent
existence
in nature to Oppenheimer\cite{oppvol}
and others, led Oppenheimer himself and
Snyder\cite{oppsny} to deduce that complete gravitational collapse
should ensue.  By putting some interior matter to collapse, matched to
a Schwarzschild exterior, it was found by them that the radius of the star
crosses its own gravitational radius and an event horizon forms with
radius $r_+$, thus discovering Schwarzschild black holes in
particular and the black hole concept in general.  Note that when
there is a star $r_+$ is the gravitational radius of the star, whereas
in vacuum $r_+$ is the horizon radius of the spacetime, so that when
the star collapses, the gravitational radius of the star gives place to
the horizon radius of the spacetime.
In its full vacuum
form, the Schwarzschild solution represents a wormhole, with its two
phases, the expanding white hole and the collapsing black hole phase,
connecting two asymptotically flat universes, see \cite{zn} \hskip -0.1cm.
There are other black holes in general relativity, belonging to the
Kerr-Newman family, having as particular cases, the
Reissner-Nordstr\"om solution with mass and electric charge, and
the Kerr solution with mass and angular momentum, see \cite{mtw} \hskip -0.1cm.
Classically,
black holes are well understood from the outside.  For their inside,
however, it is under debate whether they harbor spacetime
singularities or  have a regular core.  Clearly, the understanding
of the black hole inside is an outstanding problem in gravitational
theory.  Quantically, black holes still pose problems related to the
Hawking radiation and entropy.  Both are low energy quantum gravity
phenomena, whereas the singularity itself, if it exists, is a full
quantum gravity problem.  Black holes form quite naturally from
collapsing matter, and the uniqueness theorems are quite powerful, but
a time immemorial question is: Can there be matter objects with radius
$R$ obeying $R=r_+$?

I.e., are there black hole mimickers? Unquestionably, it is of great
interest to conjecture on the existence of maximum compact objects
that might obey $R=r_+$.  Speculations include gravastars, highly
compact boson stars, wormholes, and quasiblack holes.  Here we
advocate the quasiblack hole. It has two payoffs. First, it shows the
behavior of maximum compact objects and second, it allows a different
point of view to better understand a black hole, both the outside and
the inside stories.  To bypass the Buchdahl bound and go up to the
stronger limit $R\geq r_+$, that excludes trapped surfaces within
matter, one has to put some form of charge. Then a new world of
objects and states opens up, which have $R=r_+$.  The charge can be
electrical, or angular momentum, or other charge.  Indeed, by putting
electric charge into the gravitational system,
Andr\'easson\cite{andreasson2009} generalized the Buchdahl bound and
found that for those systems the bound is $R\geq r_+$. Thus, systems
with $R= r_+$ are indeed possible, see \cite{lzancsharp} for a
realization of this bound, and \cite{dadhich} for some physical
implications.  Systems with $R=r_+$ are called quasiblack holes.

\begin{figure}
\centering
\includegraphics[scale=0.80]{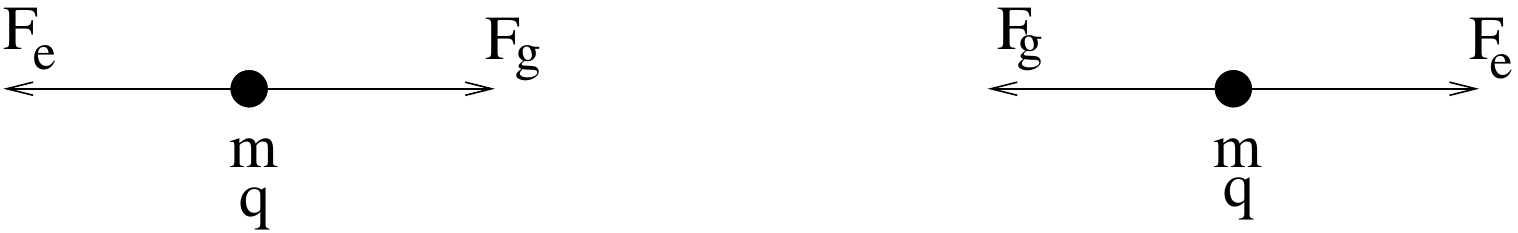}
\caption{Particles with
mass equal to charge, $m=q$, in equipoise in Newton-Coulomb
gravitation.}
\label{particlesinequipoise}
\end{figure}

To see how it works, let us start with static electrically charged
massive particles in Newtonian gravitation, i.e., let us work in
Newton-Coulomb theory.  Two massive electrically charged particles
with electric charges of the same sign, each of mass $m$ and charge
$q$ and separated by a distance $r$, attract each other
with a gravitational force given by $F_{\rm
g}=\frac{m^2}{r^2}$ and repel
each other with an electric force given by $F_{\rm
e}=\frac{q^2}{r^2}$, see Figure~\ref{particlesinequipoise}.
When $m=q$,
clearly the forces of attraction and of repulsion on each particle are
the same and one has $F_{\rm g}=F_{\rm e}$.  In this set up, a
particle at rest remains at rest in equipoise with the other particle.
Now, we can put another particle into the system with the same mass
$m$ and charge $q$, and indeed any number of these particles, even a
continuous distribution, with any symmetry, in any configuration, and
the result holds, the system stays in equipoise.
Incidentally, Mitchell
and Laplace dark stars of Newtonian gravitation, which were never implemented
as concrete systems, can easily be built from this type of matter,
by making from it an actual ball with radius $R=r_+$, such
that the escape velocity from its surface is equal to
the speed of light. 

General
relativity plus electric charged matter and electromagnetism yields
the Einstein-Maxwell system of equations.  For a static situation, one
can write the line element as
$ds^2=-W^2(x^i)\,dt^2+g_{ij}(x^k)\,dx^i\,dx^j$, with $(t,x^i)$ as the
time and spatial coordinates, respectively, and $i$ a spatial index
running from 1 to 3, $W(x^i)$ the metric potential, and $g_{ij}(x^k)$
the metric form for the 3-space.  In the case where $W^2(x^i)$ depends
strictly on the electric potential $\phi(x^i)$, i.e., $W^2=W^2(\phi)$,
then in electrovacuum the following relation holds, $W^2=
\left(\phi+b\right)^2+c$, where $b$ and $c$ are constants.  Moreover,
in the particular case that $c=0$, and so $W^2(\phi)=
\left(\phi+b\right)^2$, one can show that the solution corresponds to
two general relativistic particles, i.e., two  black holes,
with mass equal to charge $m=q$, so extremal black holes,
in equilibrium.  The solution also
holds for a number of extremal black holes scattered around in
equipoise, and generically such type of solutions is called
Majumdar-Papapetrou.  If one includes matter, maintaining the
condition $W^2(\phi)= \left(\phi+b\right)^2$, then the matter density
$\rho$ and the electric density $\rho_{\rm e}$ are related by
$\rho=\rho_{\rm e}$.  This type of matter is called extremal
matter or Majumdar-Papapetrou matter, and the corresponding
Majumdar-Papapetrou solutions are the general relativistic versions of
the simple Newtonian gravitation solutions.

\begin{figure}
\centering
\includegraphics[scale=1.60]{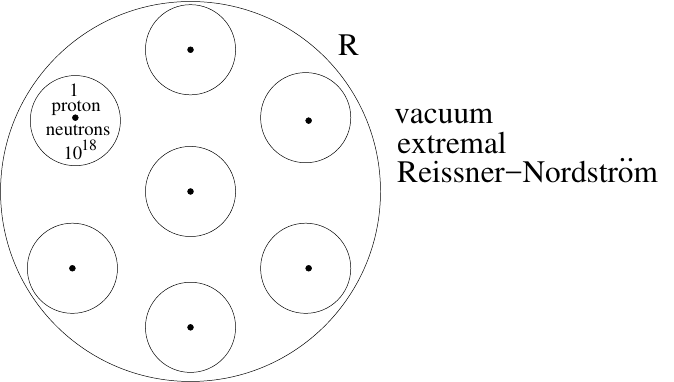}
\caption{A Bonnor star made of clouds of extremal
matter in the interior joined to a vacuum extremal
Reissner-Nordstr\"om solution.}
\label{bonnorstarofclouds}
\end{figure}

One can make a star out of Majumdar-Papapetrou, i.e., extremal, matter.
One puts a boundary at some radius $R$ on the matter, with the
interior being of extremal matter and the exterior being an extremal
Reissner-Nordstr\"om spacetime. The global solution, interior plus
exterior, is a Bonnor star\cite{bonnorstar1,bonnorstar2}.  Examples
of Bonnor stars can be given.  One example is a star made of a
continuous fluid distribution of small clouds, such that each cloud
has $10^{18}$ neutrons and 1 proton, see
Figure~\ref{bonnorstarofclouds}.  The star has, at its boundary, $m=q$
and the outside is extremal Reissner-Nordstr\"om.  Another example is
a spherical star made of any continuous distribution of
supersymmetric, or otherwise, stable particles, each particle with
mass equal to charge. The star has then, of course, total mass $m$
and total charge $q$ obeying $m=q$.

It is clear that for any star radius $R$ the  star is in
equilibrium. In a series of quasistatic steps, one can bring the
radius $R$ of the star into its own gravitational radius $r_+$, i.e.,
one can achieve a configuration for which $R=r_+$, as near as one
likes.  Since $r_+$ is the gravitational radius of the configuration,
which is on the verge of becoming a horizon, something special has to
happen. Indeed, at $R=r_+$ a quasiblack hole forms.
Thus, a quasiblack hole is an object, a star for instance,
with its boundary at its own gravitational radius.
In this sense quasiblack holes are the real general relativistic
successors of the Mitchell and Laplace stars.
Moreover, since they can be realized as stars,
and these stars are frozen at their own gravitational
radius, quasiblack holes epitomize naturally
the concept of a frozen star, a name
that Zel'dovich and Novikov gave instead
rather to mean a black hole and that was in turn superseded
by it, see \cite{zn} \hskip -0.1cm.

\newpage
\section{Examples of quasiblack holes }

\begin{figure}
\centering
\includegraphics[scale=1.05]{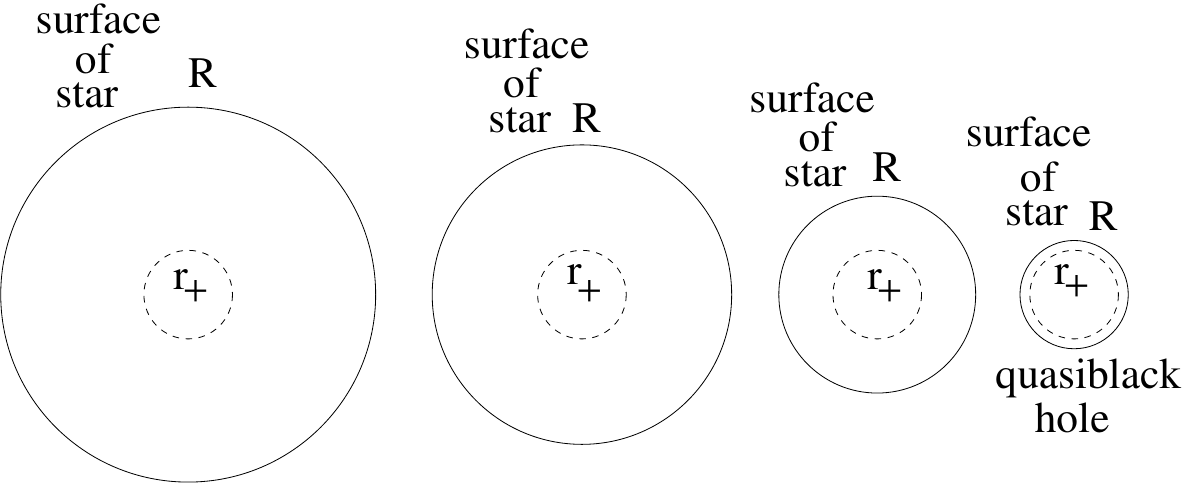}
\caption{Sequence of star configurations
to a quasiblack hole where $R=r_+$.}
\label{sequencetoquasiblackhole}
\end{figure}

Stars that can be brought to the quasiblack hole state
do not need to be Bonnor stars, these are
only one example that yields quasiblack holes.
Under certain conditions, several type of stars
that shrink to $R=r_+$, do form quasiblack holes,
see Figure~\ref{sequencetoquasiblackhole}.
For instance,
it is possible to have stars with
a nonextremal interior, that nonetheless
the condition $m=q$ at the boundary is obeyed.
These stars have as exterior the extremal Reissner-Nordstr\"om
solution, and the quasiblack hole  with
$R=r_+$ is a solution. 
Furthermore, one can generalize the concept
of a quasiblack hole to simply a solution
in which one finds $R=r_+$. 

There are many examples of quasiblack holes.  They are:
(i) Majumdar-Papapetrou quasiblack holes asymptotic to the extremal
Reissner-Nordstr\"om solution\cite{LemosWeinberg2004}.
(ii) Bonnor quasiblack holes with a sharp boundary\cite{lemoszanchinbonnorstar},
see also\cite{bonnorstar1,bonnorstar2}.
(iii) Spheroidal quasiblack holes
made of extremal charged matter\cite{Bonnor2010}.
(iv) Quasiblack holes with pressure: Relativistic charged spheres as
the frozen stars\cite{LemosZanchin2010,alz,alz2,plethora}, see also
\cite{deFeliceetal1995}.
(v) Yang-Mills-Higgs magnetic monopole quasiblack holes
\cite{LueWeinberg2000,LemosZanchin2006}.
(vi) Rotating matter at the extremal limit
\cite{BardeenWagoner1971,meinel}.
(vii) Quasiblack holes made of fundamental
fields\cite{bronnzasla}.
(viii) Matter with spin in Einstein-Cartan theory
at the quasiblack hole state, an example
that can be worked out.
(ix) Quasiblack hole shells of matter, i.e., a thin shell at its
own gravitational radius, 
with zero pressure in the extremal case and
unbound pressure in the nonextremal case,
as an exercise for a student in general relativity
and gravitation\cite{problembook}.
For a review of these examples 
see also\cite{Lemosreview}.

Since there are ubiquitous solutions one should consider
the core properties of those solutions, the most independently
as possible from the matter they are made, in much the same 
way as one does for black holes.

\section{Definition of quasiblack holes and generic properties}

Consider a static spherically symmetric
line element written in $(t,r,\theta,\phi)$
spheric coordinates as
\begin{equation}
ds^{2}=-B(r)\,dt^{2}+A(r)\,dr^{2}+r^{2}\left(d\theta
^{2}+\sin ^{2}\theta \,d\phi ^{2}\right)\,,
\label{metric1}
\end{equation}
where $A$ and $B$ are metric potentials depending on
the radial coordinate $r$,
$A=A(r)$ and $B=B(r)$.
Assume the line element represents 
an interior and an exterior region, and so 
is valid and nonsigular
for $0\leq r<\infty$. At infinity the metric is asymptotically
flat. 

Consider that the solution
for the metric potentials $B(r)$ and 
$\frac{1}{A(r)}$
has the following properties: 
(a) the function $\frac{1}{A(r)}$ attains a minimum at some
$r_{\ast}\neq 0$, such that $\frac{1}{A(r_{\ast})}=\epsilon$, with
$\epsilon \ll 1$. If one prefers an invariant definition one can
replace $\frac{1}{A(r)}$ by $(\nabla r)^{2}$, where $\nabla$ is the
gradient.
(b) For such a small but nonzero $\epsilon$ the configuration is
regular everywhere with a nonvanishing metric function $B(r)$;
(c) In the limit $\epsilon \rightarrow 0$ the metric coefficient
$B(r)\rightarrow 0$ for all $r\leq r_{\ast}$. In this limit $r_{\ast}$
is the horizon radius $r_+$, $r_{\ast}=r_+$.  These three features
define a quasiblack hole. For further details, see
\cite{qbhproperties2007,bhmimickers2008}.

\begin{figure}
\centering
\includegraphics[scale=1.60]{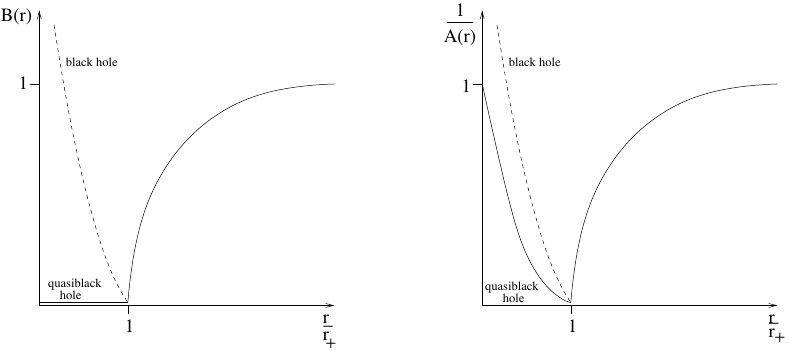}
\caption{The metric potentials $B(r)$ and $\frac{1}{A(r)}$ as
functions of $r$ for an extremal quasiblack hole and for an extremal
black hole.  In the region $r<r_+$ the functions are totally different
while in the exterior they are the same.}
\label{metricpotentialsqbh}
\end{figure}

It is relevant to
compare the form of the metric
potentials of a quasiblack hole with the 
form of the metric
potentials of an extremal electrically charged black hole.
For an extremal electrically charged black hole, i.e., an
extremal Reissner-Norsdtr\"om black hole with mass equal to charge,
$m=q$, one has for the metric potentials the relations
$B(r)=\frac{1}{A(r)}=\left(1-\frac{r_+}{r}\right)^2$, with also
$r_+=m$.
In Figure~\ref{metricpotentialsqbh}
typical plots of $B(r)$ and $\frac{1}{A(r)}$ as functions of $r$
are displayed
for quasiblack and black holes showing clearly the differences
between the two cases.

There are several important and generic properties of quasiblack holes
that should be mentioned\cite{qbhproperties2007,bhmimickers2008},
see also\cite{zaslavskii}.
1. A quasiblack hole is on the verge of forming an event horizon,
instead, a quasihorizon appears.
2. Quasiblack holes with finite stresses must be extremal to the
outside.
3. The curvature invariants of extremal and nonextremal
quasiblack hole spacetime
remain regular everywhere. 
4. A free-falling observer in a quasiblack hole spacetime
finds in its frame infinite tidal forces at the quasihorizon. This
shows some form of degeneracy, i.e., a combination of features typical
of regular and singular systems, at the quasihorizon.
5. Both in an extremal and in a nonextremal quasiblack hole
spacetime, outer and inner regions become mutually impenetrable and
disjoint. An interesting example is the Lemos-Weinberg solution, where
the interior is a Bertotti-Robinson spacetime, the quasihorizon region
is extremal Bertotti-Robinson, and the exterior is extremal
Reissner-Nordstr\"om.
6. There are infinite redshift whole 3-regions.
7. For far away observers a quasiblack hole spacetime is
indistinguishable from that of a black hole.  Quasiblack holes are
black hole mimickers.

\section{Carter-Penrose diagram
for quasiblack holes}

Carter-Penrose diagrams are a useful tool to understand the conformal
and causal structure of a spacetime.  For a spherically symmetric star
composed of vacuum and a thin shell, the corresponding Carter-Penrose
diagram is composed of the timelike origin, the Minkowski interior
bounded by the timelike radius $R$ of the thin shell star, and the
exterior with the past and future null infinities, together with the
past and future timelike infinities and the spatial infinity. For the
Carter-Penrose diagram for a quasiblack hole, made of a Minkowski
spacetime inside, a thin shell made of some matter at the boundary
$R=r_+$, and an exterior Reissner-Nordstr\"om spacetime outside
see\cite{qbh4lemoszaslacp}, where a comparison with the thin shell
star is also made.

\begin{figure}
\centering
\includegraphics[scale=1.4]{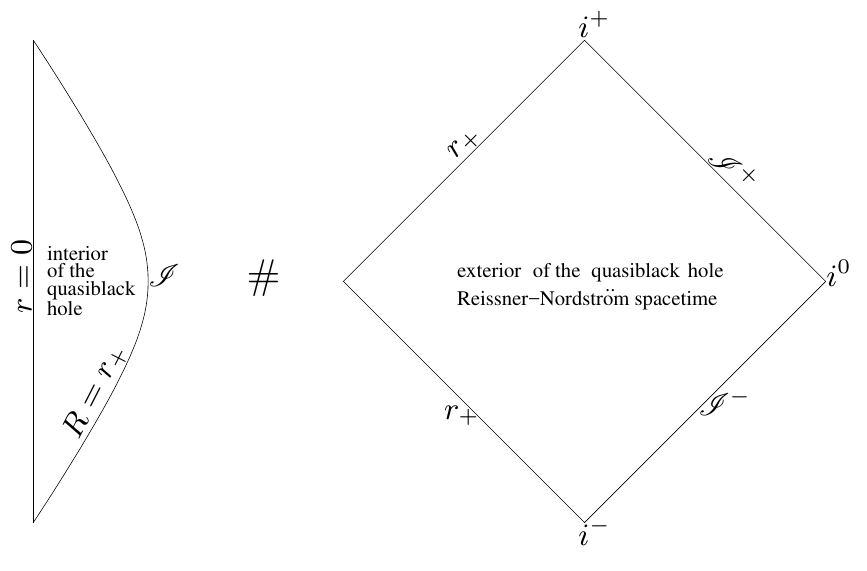}
\caption{
Carter-Penrose diagram for a generic quasiblack hole, extremal or
nonextremal.  The interior is a spacetime with matter, has the
timelike $r=0$ origin, and is bounded by a timelike line at radius
$R=r_+$ which is an impenetrable barrier that acts like an infinity
scri $\mathscr{I}$.  The exterior region is identical to the outer
Reissner-Nordstr\"om black hole region, with the past and future
horizons at $r_+$, which are null lines, the past and future null
infinities, scri minus $\mathscr{I}^-$ and scri plus $\mathscr{I}^-$,
respectively, the
past and future timelike infinities, ${i}^-$ and ${i}^+$,
respectively, and the spatial infinity ${i}^0$.  Interior and
exterior are mutually impenetrable and disjoint with $r_+$ playing the
role of a kind of singular boundary.  The symbol \# means the
connected sum of the two disjoint spacetimes.
}
\label{cp1}
\end{figure}

The Carter-Penrose diagram for a generic static spherically symmetric
quasiblack hole spacetime, extremal or nonextremal, is shown in
Figure~\ref{cp1}. There are two separated causal regions which are
mutually impenetrable and disjoint. This is connected to the fact that
tidal forces are infinite at the boundary which functions as a
barrier.  Infinite energy particles could, in principle, penetrate
this barrier, but those would destroy the spacetime.  The interior
region is composed of some matter that extends up to the timelike
boundary $R=r_+$, which acts as an infinity scri $\mathscr{I}$.
Outgoing timelike and null geodesics are reflected at the timelike
boundary $\mathscr{I}$, to become ingoing timelike and null geodesics,
respectively, and so forth.  The outer region is indistinguishable
from a Reissner-Nordstr\"om black hole outer region, it has the past
and future null horizons $r_+$, the past and future null infinities,
scri minus $\mathscr{I}^-$ and scri plus $\mathscr{I}^+$,
respectively, together with the past and future timelike infinities
and the spatial infinity, ${i}^-$, ${i}^+$ and ${i}^0$,
respectively. The null horizons $r_+$ are naked horizons as they show
some form of singular behavior.  Quasiblack holes have special causal
properties, and their Carter-Penrose diagrams show that they are a
natural blend of stars, regular black holes, and null naked horizons.

\section{Pressure properties of quasiblack holes}

A quasiblack hole has matter in its interior and so it is important to
study its pressure properties, particularly at the boundary $R=r_+$
\cite{lemoszaslavskiiqbhswithpressure}.  We restrict to the quasiblack
holes that are extremal on the exterior, the quasiblack holes
that are nonextremal have unbound pressure at $R=r_+$.

Let us analyze the case in which the matter is nonextremal in the
interior region, to start.  Imposing finite Riemann tensor components
$R_{abcd}$ in an orthonormal frame,
and denoting radial pressure by $p_r$, 
we
find $p_{r}^{\mathrm{in}}(r_{+})=-\frac{1}{8\pi
r_{+}^{2}}$.  For an extremal exterior region one also has
$p_{r}^{\mathrm{out}}(r_{+})=-\frac{1}{8\pi r_{+}^{2}}$.  So, we find
\begin{equation}
p_{r}^{\mathrm{in}}(r_{+})=p_{r}^{\mathrm{out}}(r_{+})\,,
\end{equation}
i.e., this type of quasiblack holes have continuous
pressure at the boundary, surely a neat result.  If matter in the
interior region is not electrical then
$p_{r}^{\mathrm{in}}(r_{+})=p_{r}^{\mathrm{in\; matter}}(r_{+})$, and
the quasiblack holes in question are supported by tension.
If matter
in the interior region is electrical then through the whole
interior the pressure $p_{r}^{\mathrm{in}}(r)$ is composed of a matter
part $p_{r}^{\mathrm{in\;matter}}(r)$ and of an electromagnetic part
$p_{r}^{\mathrm{in\;em}}(r)$, such that
$p_{r}^{\mathrm{in}}(r)=p_{r}^{\mathrm{in\;matter}}(r)+
p_{r}^{\mathrm{in\;em}}(r)$, but at $r_+$ one finds obligatory that
$p_{r}^{\mathrm{in\;matter}}(r_{+})=0$, and so
$p_{r}^{\mathrm{in}}(r_{+})=p_{r}^{\mathrm{in\;em}}(r_{+})$.  This
means that in this case all pressure support at the boundary comes
from the electric part and there is no pressure support from matter
at $r_+$. This case, in contrast to the previous one, can arise
from a quasistatic
shrinking star. Indeed, a star with radius $R$ has to have
zero pressure at the boundary, $p_{r}^{\mathrm{in\;matter}}(R)=0$,
so that the boundary
is static. One can then contract the star in quasistatic
way to $r_+$ to yield a quasiblack hole with 
$p_{r}^{\mathrm{in\;matter}}(r_+)=0$. 
Note also, that since the matter inside is not
extremal, by assumption, there is a jump 
in the density at $r_+$, but such jumps
in density
pose no
problems.

Let us now analyze the quasiblack hole  case in which the matter is
extremal in the
interior region. 
Imposing finite Riemann tensor components $R_{abcd}$
in an orthonormal frame we
find 
\begin{equation}
p_{r}(r_{+})=-\rho (r_{+})\,. 
\end{equation}
This is the same condition as
found for dirty black holes.
One can 
prove further, in this case of matter being
extremal in the
interior region, the following:
(i) One cannot build an interior
extremal quasiblack hole entirely from
phantom matter, i.e., one cannot build
such a quasiblack hole
with the matter violating the null energy condition,
namely,  $p_{r}+\rho
<0$, everywhere inside. 
(ii) In case there is phantom matter, it cannot border the
quasihorizon, it has to be
in the inner region. At least in a vicinity of the
quasihorizon the null energy condition, $p_{r}+\rho \geq 0$,
is satisfied
\cite{lemoszaslavskiiqbhswithpressure}.

\section{Mass formula for quasiblack holes}

The quasiblack hole, being an object on the verge of
becoming a black hole shares several properties
with black holes themselves.
We now work out the mass formula
for quasiblack holes.
One needs two steps. In the first one makes sure
that at the quasihorizon the Kretschmann scalar
is finite. In the second one uses the Tolman mass definition
to find a mass formula for quasiblack holes.

For the first step it is useful
to rewrite the metric given in Eq.~(\ref{metric1})
and put it 
in Gaussian coordinates
valid near the quasihorizon.
Redefining the metric potentials
$B$ and $A$ in Eq.~(\ref{metric1})
such that $N=B$ and the new radial coordinate
$l$ is $dl=\frac{dr}{\sqrt A}$, the metric
takes the form
\begin{equation}
ds^{2}=-N^{2}dt^{2}+dl^{2}+g_{ab}dx^{a}dx^{b}\,,
\label{metric2}
\end{equation}
with $a,b=1,2$.
The Kretschmann scalar $\mathrm{Kr}$
for this metric is  given by
\begin{equation}
\mathrm{Kr}=P_{ijkl}P^{ijkl}+4C_{ij}C^{ij}\,,
\end{equation}
where  $i,j=1,2,3$ are spatial indices, 
$P_{ijkl}$ is the curvature tensor for
a $t=\mathrm{const}$ hypersurface, and
\begin{equation}
C_{ij}=\frac{N_{; ij}}{N}\,,
\label{cij}
\end{equation}
with
$\,_{;i}$ denoting a covariant derivative. As the metric of the
3-space is positive definite, all terms enter the
expression with a positive sign, so
if we impose finiteness for 
$\mathrm{Kr}$ this means 
each term should
be finite separately. The scalar $P_{ijkl}P^{ijkl}$
for the metric~(\ref{metric2}) is clearly
finite. So we have to deal with the term $4C_{ij}C^{ij}$.
By definition, quasiblack hole implies
that the metric potential $N$ in  Eq.~(\ref{metric2})
satisfies $N=N(x^a)\to0$.
Choose $l=0$ on the surface
of the object, without loss of generality. Putting
$\,^{\prime }\equiv \frac{\partial }{\partial l}$ and
$\,_{;a}$ the covariant derivative for
the two-metric $g_{ab}$ in  Eq.~(\ref{metric2}), we find
from Eq.~(\ref{cij})
\begin{equation}
\lim_{l\rightarrow 0}C_{ll}=\lim_{l\rightarrow 0}\frac{N^{\prime
\prime }}{ N_{0}}\,, \quad
\lim_{l\rightarrow 0}C_{al}=\lim_{l\rightarrow 0}\frac{N_{;a}^{\prime
}}{ N_{0}}.
\label{fkr}
\end{equation}
Finiteness of
Kr
implies
$\lim_{\epsilon \rightarrow 0}\lim_{l\rightarrow 0}
N^{\prime \prime }=0$ and $\lim_{\epsilon \rightarrow 0}
\lim_{l\rightarrow 0}N_{;a}^{\prime }=0$.
Near the quasihorizon, we expand the metric function $N$ 
as
$N=N_{\mathrm{0}}+$ $\kappa _{1}(x^{a},\epsilon )l+\kappa
_{2}(x^{a},\epsilon )\frac{l^{2}}{2!}+\kappa
_{3}(x^{a},\epsilon)
\frac{l^{3}}{3!}+O(l^{4})$,
where $N_{\mathrm{0}}$ is some constant,
and  $\kappa_1$,
$\kappa_2$,
and $\kappa_3$ are functions that have to be determined.
The first part of Eq.~(\ref{fkr}) implies
$\lim_{\epsilon \rightarrow 0}\kappa _{2}=0$.
The second part of Eq.~(\ref{fkr}) implies
$\lim_{\epsilon \rightarrow 0}\kappa_{1}(x^{a},\epsilon
)=\kappa$, where $\kappa$ is a constant, which is identified with the
surface
gravity of the corresponding surface.
So at the quasihorizon one has
\begin{equation}
N=N_{\mathrm{0}}+\kappa l+\kappa_
{3}(x^{a})\frac{l^{3}}{3!}+O(l^{4})\,.
\label{nexp}
\end{equation}

For the second step we use the fact that
when there is matter there is mass
and that mass is given by the Tolman
formula. The Tolman formula
for the mass $m$ of an object is
\begin{equation}
m=\int (-T_{0}^{0}+T_{i}^{i})\sqrt{-g}\,d^{3}x\,,
\end{equation}
where $T_{0}^{0}$ and $T_{i}^{i}$ are the
components of the energy-momentum tensor,
$g$ is the determinant of the metric,
and the integral is performed over the region
of interest.
Here, it is convenient
to split the mass $m$  into three parts, namely,
\begin{equation}
m=M_{\mathrm{in}}+M_{\mathrm{surf}}+M_{\mathrm{out}}
\,,
\end{equation}
where 
$M_{\mathrm{in}}$
is the interior mass,
$M_{\mathrm{surf}}$ is the surface mass, and 
$M_{\mathrm{out}}$ is the outer mass. Let us analyze each one
in turn.
The interior mass is 
$M_{\mathrm{in}}=\int_{\mathrm{in}}(-T_{0}^{0}+T_{i}^{i})\,N\sqrt{g_{3}}
d^{3}x$, so that
$M_{\mathrm{in}}\leq N_{\mathrm{B}}\,(M_{0}+M_{k})$,
with $N_\mathrm{B}$ being the value of the metric potential
at the boundary and $M_0$ and $M_k$ being the components of the
mass in an obvious notation. 
Since for a quasiblack hole  $N_\mathrm{B}\to0$ one has
\begin{equation}
M_{\mathrm{in}}= 0
\,.
\end{equation}
For the surface, from Dirac-$\delta$
contributions, we define  $S_{\mu }^{\nu}$ as
$S_{\mu }^{\nu}=\int T_{\mu}^{\nu }\,dl$.
So that one gets
$M_{\mathrm{surf}}=\int (-S_{0}^{0}+S_{a}^{a}) \,N\,d\sigma$.
Now, one has  $8\pi S_{\mu }^{\nu }=\left([K_{\mu
}^{\nu }]-\delta_{\mu }^{\nu }[K]\right)$, where $K_{\mu }^{\nu}$
is the extrinsic curvature tensor, $[...]=[(...)_{+}-(...)_{-}]$,
and $+$ and
$-$ refer to the outer and inner
sides. After calculating $K_{\mu }^{\nu}$
for the metric~(\ref{metric2}) one
finds $M_{\mathrm{surf}}=\frac{1}{4\pi
}\int_{\mathrm{surf}}\left[ \left(\frac{ \partial N}{\partial
l}\right) _{+}- \left( \frac{\partial N}{\partial l} \right)
_{-}\right] d\sigma$, where
$d\sigma$ is the two-surface element. Now $(\frac{\partial N}{\partial
l})_{-}\rightarrow 0$, and since from Eq.~(\ref{nexp}) one has 
$N=N_{\mathrm{0}}+\kappa l+\kappa_{3}(x^{a})
\frac{l^{3}}{3!}+O(l^{4})$, one finds $(\frac{\partial
N}{\partial l})_{+}=\kappa$. So finally,
\begin{equation}
M_{\mathrm{surf}}=\frac{\kappa A_+}{4\pi }\,,
\end{equation}
$A_ +$ being the 
quasihorizon  area.
For the outer mass $M_{\mathrm{out}}$ one has
$M_{\mathrm{out}}=
\int_{\mathrm{out}}(-T_{0}^{0}+T_{k}^{k})\,N\sqrt{g_{3}}
d^{3}x$, so when one includes an electromagnetic field
one finds 
\begin{equation}
M_{\mathrm{out}}=\varphi_+q+
M_{\;\mathrm{out}}^{\mathrm{matter}}\,,
\end{equation}
where $\varphi_+$ is the electric potential
at the quasihorizon $r_+$, $q$ is the
outer electric charge, and 
$M_{\;\mathrm{out}}^{\mathrm{matter}}$ is the
matter that lingers outside $r_+$, e.g., matter in
an accretion disk.

With steps one and two in hand, one can now put all the masses
together to obtain the total mass $m$ of a static system containing a
quasiblack hole, namely, \cite{lemoszaslamass}
\begin{equation}
m=\frac{\kappa A_+}{4\pi}+
\varphi_+q+M_{\;\mathrm{out} }^{\mathrm{matter}}\,.
\label{massfor}
\end{equation}
In vacuum, $M_{\;\mathrm{out}}=0$, and the mass formula becomes $
m=\frac{\kappa A_+}{4\pi}+
\varphi_+q$.
For the extremal case $\frac{\kappa
A_+}{4\pi}$ goes to zero, as $\kappa=0$,
and since one can set $\varphi_+=1$
one gets
$m=q$ as the mass formula for a pure extremal
quasiblack hole.
Note that in the
nonextremal case the $M_{\mathrm{surf}}\neq0$ contribution
comes from $|S^\mu_\nu|\to\infty$, whereas in the
extremal case one has
$M_{\mathrm{surf}}=0$ and the
surface of the quasiblack hole makes no contribution
to the mass.
When there is rotation the quasiblack hole has angular
velocity $\omega_+$ and so there is
also angular momentum
$J$. In this case the mass formula for quasiblack holes is
\cite{lemoszaslamassangular}
\begin{equation}
m=\frac{\kappa A_+}{4\pi}+2\omega_+J+
\varphi_+q+M_{\;\mathrm{out}}^{\mathrm{matter}}\,. 
\label{massangularfor}
\end{equation}
In vacuum, $M_{\;\mathrm{out}}=0$, and the mass formula becomes $
m=\frac{\kappa A_+}{4\pi}+2\omega_+J+
\varphi_+q$.

Two comments are in order. First,
the mass formulas that appear in
Eqs.~(\ref{massfor}) and~(\ref{massangularfor}) have the same form as
the mass formulas for
pure black holes, i.e., the Smarr formula
\cite{smarr}, and for  black holes
and surroundings\cite{bch}, but
were obtained from totally different means.
For black holes, since these are vacuum solutions,
the appropriate mass definition is the Komar mass,
which is a totally different definition
from the Tolman mass that we used for quasiblack holes.
Strikingly, either mass definition yields
the same mass formula.
Second, the surface gravity $\kappa$ that
appears 
in the mass formulas of 
Eqs.~(\ref{massfor})
and~(\ref{massangularfor}) have
for black holes the interpretation
of being the acceleration a test particle
experiences at $r_+$ redshifted to infinity.
For quasiblack holes
one can  have an alternative interpretation,  
$\kappa$ being the surface density
of the matter at the surface of the
quasiblack hole. This interpretation was
unveiled when studying the mass formula
for black holes from the
membrane paradigm perspective\cite{lemoszaslamembranemass}.

\section{Entropy of quasiblack holes}

\subsection{Rationale}

A fundamental test for the theory of quasiblack holes is to find their
entropy $S$.  Imagine a collapsing body.  Its matter has some entropy that
might grow or not during the collapse, but when the surface is at the
horizon $r_+$ there is no apparent reason for the entropy to turn into
$S=\frac14 A_+$, it appears as a jump.  We will see that by working
with a quasistatic contraction and a quasiblack hole approach one sheds
some light on the origin of this $S=\frac14 A_+$ entropy.

We use the first law of thermodynamics for the matter and
gravitational fields.  In general, to find the entropy of a system one
needs an equation of state.  One then integrates the first law on a
path along the energy first, say, and then along a volume, or a
length, or some other useful quantity.  With the quasiblack hole
approach one can do differently, dispensing altogether an equation of
state for the matter.  For a quasiblack hole one picks a different
path, i.e., one chooses a sequence of configurations such that all
members remain on the threshold of horizon formation and then one
integrates over this subset of members. The answer has to be
independent of the model and independent of the equation of state for
the matter. The approach explores the fact that the boundary of the
matter tends to a quasihorizon in the quasiblack hole limit.

\subsection{Entropy of spherical quasiblack holes}

\subsubsection{Generics}

Let us suppose that there is a spherically symmetric star spacetime
composed of some interior spacetime with a fluid with energy density
$\rho$, tangential pressure, radial pressure $p_r$, and a boundary
surface at radius $R$, plus an outside Schwarzschild spacetime which
is characterized by the spacetime mass $m$ or equivalently by the
gravitational radius $r_+=2m$.  We want to take the radius of the star
to the quasiblack hole limit, $R=r_+$, analyze the first law of
thermodynamics for the system and find its entropy $S$.

For this purpose we rewrite the metric of
Eq.~(\ref{metric1}),
now putting $B(r)=N^2(r)\mathrm{e}^{2\psi (r)}$
and $A(r)=\frac{1}{N^2(r)}$, so
that 
the star's static spherically symmetric
metric is written as
\begin{equation}
ds^{2}=-N^2(r)\mathrm{e}^{2\psi (r)}dt^{2}+\frac{
dr^{2}}{N^2(r)}+r^{2}(d\theta ^{2}+\sin ^{2}\theta d\phi ^{2})\,.
\label{metricgeneral1}
\end{equation}
Einstein field equations yield for this metric 
\begin{equation}
N^2(r)=1-
\frac{8\pi}{r}
\int_{0}^{r}d\bar{r}\,\bar{r}^{2}\rho\,,
\end{equation}
\begin{equation}
\psi (r)=4\pi \int_{R}^{r}d\bar{r}\,
\frac{(\rho +p_r)\,\bar{r}}{N^2(\bar r)}
\,.
\end{equation}
There is another equation involving the
tangential pressure that we do not write
here.
The
matter is constrained to the region $r\leq R$,
so that for $r\geq R$ one has $\psi(r)=0$
and $N^2=1-\frac{r_+}{r}$, i.e., the Schwarzschild solution.

\begin{figure}
\centering
\includegraphics[scale=0.38]{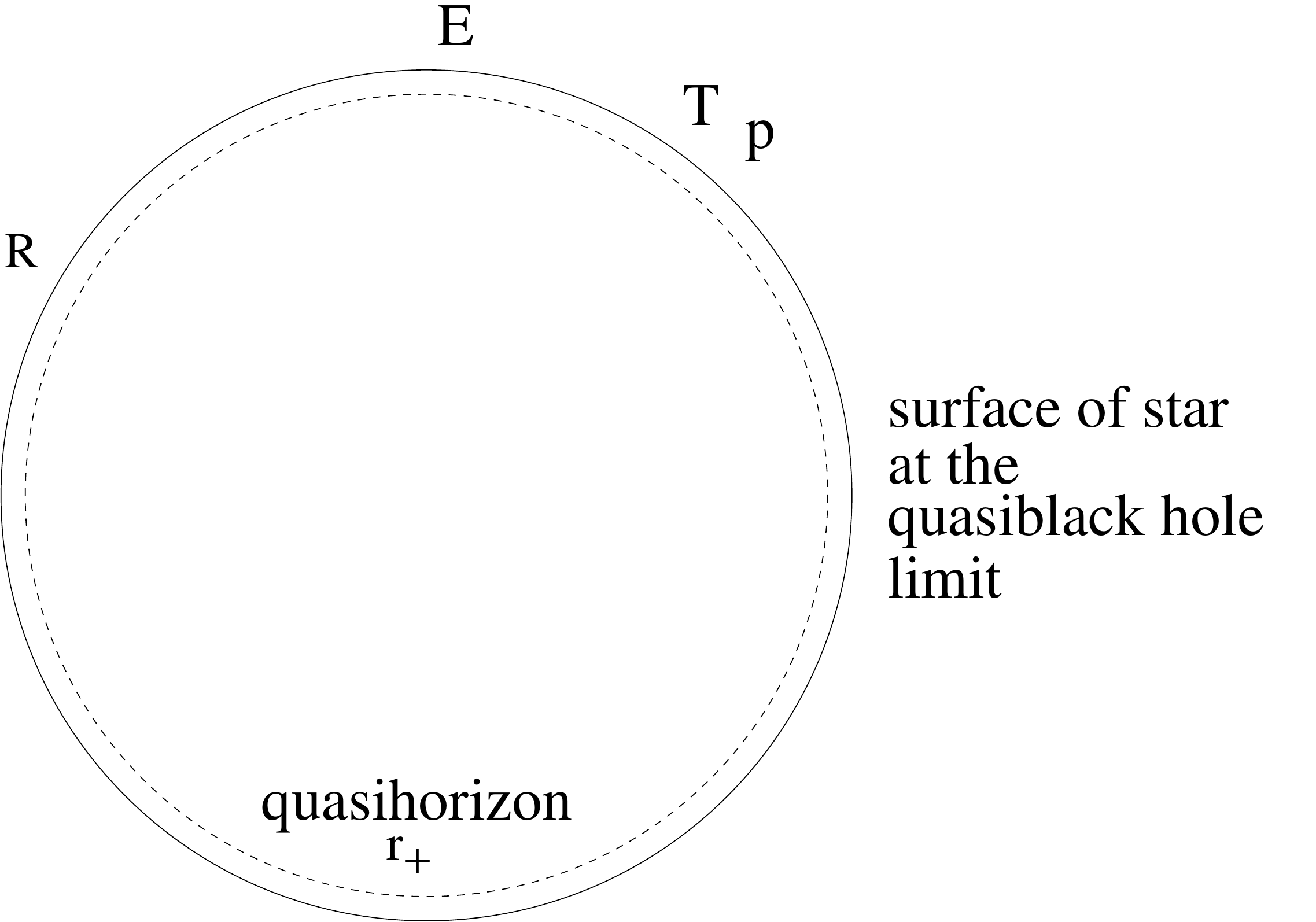}
\caption{A spherical star approaching its quasiblack hole state
and its thermodynamic quantities at
the quasihorizon $R=r_+$.}
\label{1stlawspherical}
\end{figure}

For a gravitational system
the first 
law of thermodynamics
is given in terms of
boundary values
where the boundary
can be put at any radius $r$
\cite{by}. 
Since
we are working with a spherically
symmetric spacetime the first 
law of thermodynamics  can be written as
\begin{equation}
TdS=dE+p\,dA\,,
\label{1stlaw11}
\end{equation}
where $T$ is the local temperature, $S$
the entropy, $E$ the energy, $p$ the tangential
pressure, and $A$ the area,
with all these quantities being
locally defined quantities
at a radius $r$. In particular
the quantities are
well defined
at the interface $R$, see Figure~\ref{1stlawspherical}.

We want to find the entropy $S$ when the radius of the star is at its
own gravitational radius, $R=r_+$, i.e., at the quasiblack hole limit.
We use the quasiblack hole condition that $N\to0$. Then, we change
simultaneously the radius $R$ and the
thermodynamic energy $E$, keeping the
interface $R$ near the gravitational radius $r_+$, with the condition
$N\to0$, for all configurations of interest in this process.
Mathematically we write $R=r_+(1+\delta)$ for some $\delta$ and send
$\delta \rightarrow 0$ ensuring the star is kept near the
quasihorizon. Then we integrate the first law along such a sequence of
quasihorizon configurations, counting different members of the same
family of states and obtain $S$ at a given $r_+$.

In the calculations to be performed
there is an assumption that should be mentioned.
The local
temperature $T$  and the temperature at infinity $T_0$,
say, are related by the Tolman formula, namely, $T=\frac{T_0}{N}$,
where in the quasiblack hole state $N=\sqrt\delta$.  But
near the quasihorizon $R=r_+$, the backreaction of quantum
fields is divergent unless $T_0$ is the Hawking temperature $T_{\mathrm{H}}$,
so that at the quasiblack hole limit one has to put
 $T_0=T_{\mathrm{H}}$.

\subsubsection{Example: Spherically symmetric quasiblack holes
with vacuum Minkowski interior
and a thin shell at the boundary}

An interesting system to work with, which allows not only a
thermodynamic solution
through the quasiblack hole approach but also an exact
thermodynamic solution, is a thin shell system.
The whole system is composed of
Minkowski spacetime in the interior,
a thin shell of radius $R$ at the junction, and
Schwarzschild spacetime in the exterior
region.
The exterior
spacetime mass is $m$ and its
gravitational radius is $r_+=2m$,
so that $m$ and $r_+$ can be interchanged.
The metric potential $N$
of Eq.~(\ref{metricgeneral1}) for the
exterior is given by
\begin{equation}
N^2=1-\frac{r_+}{R}
\,,
\end{equation}
and the other metric potential is $\psi=0$. 
The thin shell at radius $R$
has proper mass $M$ which is
also the thermodynamic energy $E$\cite{by},
i.e., $M=E$.
The relation between $m$, $E$ and $R$ is 
$m=E-
\frac{E^{2}}{2R\,\,\,\,\,}$, and since
$r_+=2m$, one has
$r_+=2E-
\frac{E^{2}}{R\,\,\,\,\,}$, or solving for $E$ gives
the proper mass or thermodynamic energy $E$ as
\begin{equation}
E=R\left(1-N\right)\,.
\label{massE}
\end{equation}
The tangential pressure $p$
that supports the thin shell
is taken from the junction condition
and is given by 
\begin{equation}
p=\frac{\left( 1-N\right)^2}{16\pi R\,N}
\,.
\label{pshell}
\end{equation}
The shell has a temperature
$T$ and area $A=4\pi R^2$. 
The first law is given in
Eq.~(\ref{1stlaw11}).

Now we use the quasiblack hole approach.  In general to integrate the
first law~(\ref{1stlaw11}) one needs equations of state for $p$ and
$T$.  But not here. Here we want to integrate the first law when the
system is near $r_+$.  Then, in the process of integrating the first
law, all three quantities $R$, $r_+$, and $E$, change, but we impose
that they change in such a way that $R=r_{+}(1+\delta)$, and so
$N^2=\delta$, with $\delta$ small.  In other words, we change
simultaneously and proportionally $R$ and $r_+$ when passing from one
equilibrium configuration to another.  Then the pressure term $p$ in
Eq.~(\ref{pshell}), $p= \frac{1}{16\pi R\,N} \approx \frac{1}{16\pi
\,r_+\,N} $, is huge since $N=\sqrt\delta$ is small. Also one has
$dA=8\pi r_+ dr_+$.  The term $dE$ given from Eq.~(\ref{massE}) is
$dE\approx dR\approx dr_{+}$, and so is negligible.  The local
temperature $T$ at the shell and the temperature at infinity $T_0$,
say, are related by the Tolman formula, namely, $T=\frac{T_0}{N}$,
where in the quasiblack hole state $N=\sqrt\delta$.  One should now
note that near the quasihorizon $R=r_+$ the backreaction of quantum
fields is divergent unless $T_0$ is the Hawking temperature
$T_{\mathrm{H}}$, i.e.,
$T_0
=T_{\mathrm{H}}=\frac{\kappa }{2\pi}=\frac{1}{4\pi r_+}$.  Then,
putting altogether in the first law~(\ref{1stlaw11}), we find $dS=2\pi
dr_+$, i.e., 
\begin{equation}
S=\frac14 A_+
\,,
\end{equation}
where $A_+=4\pi r_+^2$ is the area of the quasihorizon,
and we have put the constant of integration
to zero.  This is the
Bekenstein-Hawking entropy for black holes.
For details see \cite{lemoszaslaentropyquasibhs1}.
For a related treatment involving the
membrane paradigm see\cite{lemoszaslamembraneentropy}.

Now we use the thin shell approach, as the thin shell spacetime offers
an alternative route to perform the calculations different from the
quasiblack hole approach, and permits an exact solution of the
thermodynamic problem.  Indeed, the solution for the thin shell
thermodynamic problem is exact and valid for all $R$, not only $r_+$.
Take the first law as given in Eq.~(\ref{1stlaw11}).  Again, the
junction condition gives the tangential pressure for any radius $R$ as
$p=\frac{1}{16\pi}\frac{(1-N)^2}{R\,N}$, with $N^2=1-\frac{r_+}{R}$.
Now, take into account the integrability condition that comes out of
the first law~(\ref{1stlaw11}), and change variables from $(E,R)$ to
$(r_+,R)$.  This integrability condition gives that the local temperature
$T$ has to have the form $T=\frac{T_{0}(r_+)}{N}$, necessarily, where
$T_{0}(r_+)$ has the usual meaning of the temperature measured by an
observer at infinity. Clearly this equation for $T$ is the other
equation of state needed to integrate the first law~(\ref{1stlaw11}).
Inserting the two equations of state, namely for $p$ and for $T$, into
the first law we find $dS=\frac{dr_+}{2T_{0}(r_+)}$.  Hence the
entropy can be found by direct integration once the equation of state
$T_{0}(r_+)$ is known.  In particular, we can choose then $T_{0}$ as
the Hawking temperature $T_{\rm
H}$, i.e.,   $T_{0}(r_+)=T_{\rm
H}=\frac{1}{4\pi r_+}$.  In this case we find $S=\frac14 A_+$. For a
thin shell with the Hawking temperature, the formula is valid for any
shell radius $R$, including the quasiblack hole radius $R=r_+$.
Moreover, for a quasiblack hole $T_{0}(r_+)=T_{\rm
H}=\frac{1}{4\pi r_+}$ is the only equation of state possible,
to avoid infinite back reaction effects. 
so $S=\frac14 A_+$ is the only solution for the entropy in
this case, in conformity
with the quasiblack hole approach
of the last paragraph. Note also that the entropy of a thin shell
does not depend on $R$. This is a consequence of the fact that 
there is no matter inside, 
one has $\frac{ \partial S}{\partial R}=0$ everywhere.
For the thermodynamics of a Schwarzschild thin shell
see \cite{martinez}, the thermodynamics
of the thin shell at its gravitational radius limit
and quasiblack hole
limit were taken in this paragraph.

\subsubsection{Example: Spherically symmetric quasiblack holes with
a continuous distribution of matter in the inside}

We now find the entropy of a spherically symmetric quasiblack hole
with a continuous distribution of matter in the inside, rather than a
thin shell.  For spherical symmetric general configurations with a
continuous distribution one has that $\frac{\partial S}{\partial R}$
is nonzero.  In general, an exact thermodynamic solution for a star
with a continuous distribution of matter cannot be found. For such a
star, the temperature depends on $r_+$ and $R$ in some complicated
fashion, so that the entropy $S$ of the star will also depend on $r_+$
and $R$. The integrability conditions are of no use and it is virtually
impossible to make progress. However, if the star is at the
quasiblack hole limit, it is possible
to implement the quasiblack hole approach and find the entropy
from the first law of thermodynamics.

Consider then a general metric for a spherically symmetric
distribution of matter. The metric is the one given in
Eq.~(\ref{metricgeneral1}). For spherically symmetric systems, the
first law of thermodynamics given in Eq.~(\ref{1stlaw11}) can be
written alternatively and usefully in terms of the temperature at
infinity $T_0$, $r_+$, $R$, and the radial pressure of the matter as
\begin{equation}
T_{0}\,dS=\frac12\exp {\psi (R)}\,
\left( dr_++8\pi p_{r}R^{2}dR\right) \,.
\label{1stlawawk}
\end{equation}
Although, without knowing the system details, one cannot find
$S(r_+,R)$ in general, for spherical quasiblack holes one can bypass
this lack of knowledge and deduce the entropy of the system through a
series of steps.  The steps are: (i) Since we want $R\rightarrow
r_{+}$ we also have to put $T_{0}$ as the Hawking temperature in order
that the backreaction of quantum fields remains finite,
$T_{0}\rightarrow T_{\mathrm{H}}$. Now, for the
metric~(\ref{metricgeneral1}), $T_{\mathrm{H}}$ is given by
$T_{\mathrm{H}}=\frac{\mathrm{e}^{\psi (r_{+})}}{4\pi }\frac{
dN^2(r)}{dr}(r_{+})$, i.e., at the quasihorizon $R=r_+$ one finds, $
T_{0}=T_{\mathrm{H}}=\frac{\mathrm{e}^{\psi
(r_{+})}}{4\pi\,r_{+}}\left( 1-8\pi \rho (r_{+})r_{+}^{2}\right) \,.
$ Thus, substituting into the first law
of Eq.~(\ref{1stlawawk})
we obtain $dS=$ $\exp (\psi
(R)-\psi ({r_{+}}))\, 2\pi r_+\frac{dr_++8\pi p_{r}R^2dR}{1-8\pi \rho
_{+}r_{+}^{2}}$.  This equation gives the change of the entropy in
terms of the changes of $r_+$ and $R$.  (ii) We are interested in the
quasihorizon limit, we want to move along the line $R\approx r_{+}$ in
the space of parameters, so that $dR\approx dr_{+}$.  Then for $
R\rightarrow r_{+}$, the factor $\exp (\psi (R)-\psi ({r_{+}}))$ in
$dS$ drops out, so $dS= 2\pi r_+\frac{1+8\pi p_{r+}r_+^2}{1-8\pi
\rho_{+}r_{+}^{2}}dr_+$.  Also, from the regularity conditions on
the pressure at the quasihorizon one has $p_{r}(r_{+})=-\rho (r_{+})$
and the terms in the numerator and denominator cancel.  Thus, putting
this into the first law yields $dS=2\pi r_{+}dr_{+}$.  (iii) Then upon
integration on finds $S=\frac{1}{4}\,A_+$, i.e., one gets back the
Bekenstein-Hawking entropy.

\subsection{Entropy of nonspherical quasiblack holes}

\begin{figure}
\centering
\includegraphics[scale=0.38]{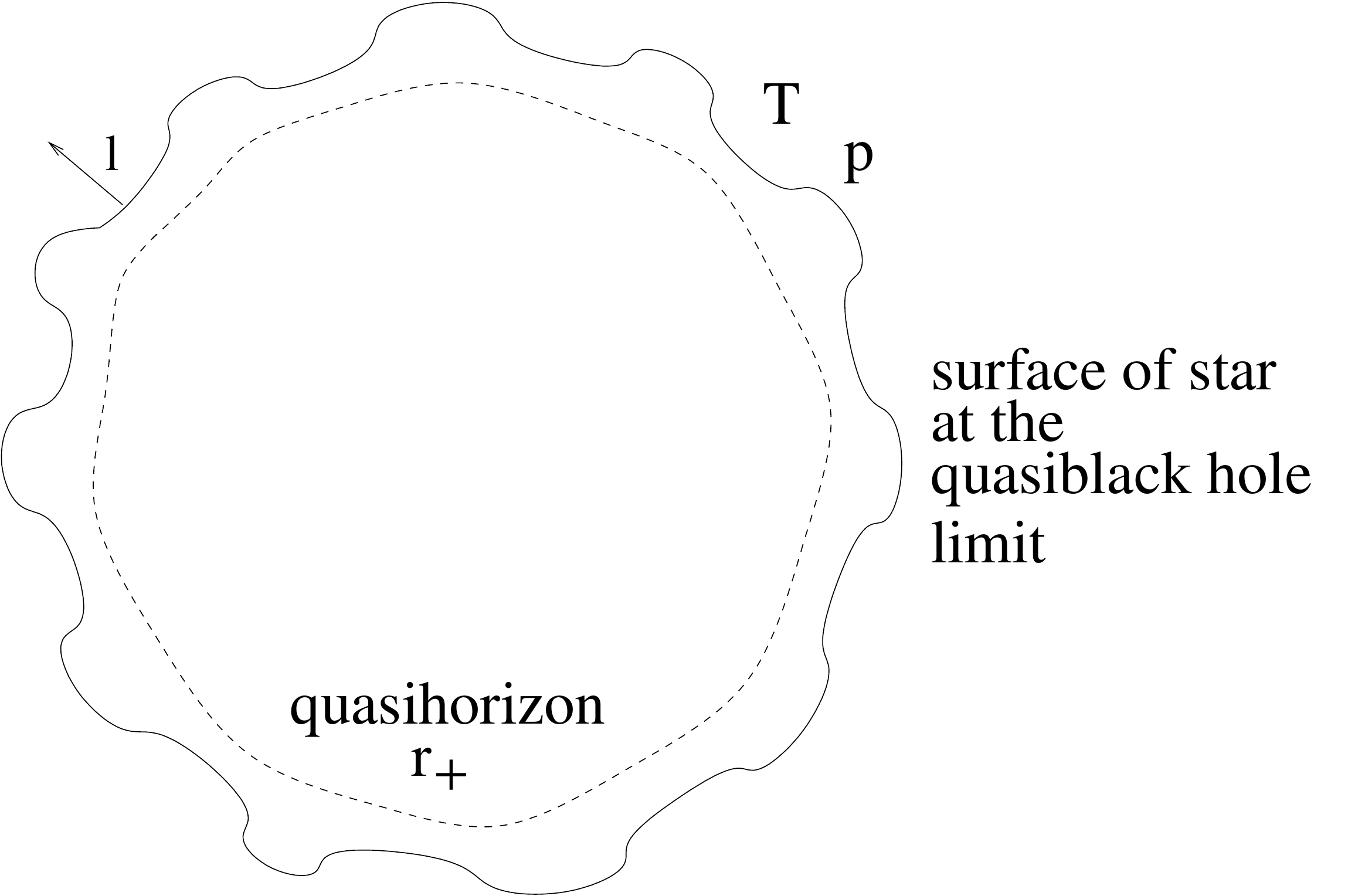}
\caption{A nonspherical star approaching its
quasiblack hole state and its thermodynamic quantities at
the quasihorizon.}
\label{nonsphericalqbh}
\end{figure}

One can also study the thermodynamics and the entropy of nonspherical
quasiblack holes, see Figure~\ref{nonsphericalqbh}.  In this generic
nonspherical case it is appropriate to use again Gaussian coordinates
for which the line element can be written as in Eq.~(\ref{metric2}).
Suppose that the boundary of the compact body is at $l=0$ without loss
of generality.  

The local Tolman temperature at the surface $l=0$ is
denoted by $T$, whereas $T_{0}$ is the temperature at asymptotically
flat infinity. The relation between the two temperatures is the Tolman
relation $T=\frac{T_{0}}{N}$.
Since now there is no spherical symmetry
we cannot write the first law,
Eq.~(\ref{1stlaw11}) in terms of the
quantities $S$, $E$, and $A$, 
which were defined over the whole
sphere. 
Instead we have to resort to
densities, i.e., quantities per unit area. They are
$s$, $\varepsilon$, and $a$, the entropy density, energy
density,  and unit area, respectively.
The first
law of thermodynamics in terms of these
densities at the boundary can then be written as\cite{by}
\begin{equation}
Td(\sqrt{g}\,s)=d(\sqrt{g}\,{\varepsilon})+
\frac{\Theta ^{ab}}{2}\sqrt{g}
\,dg_{ab}\,.
\label{1stlawdensity}
\end{equation}
Here $g$ is the determinant of the two-metric $g_{ab}$
defined in Eq.~(\ref{metric2}), 
the  energy density on the layer
is defined as 
${\varepsilon}=\frac{K}{8\pi}$
with 
$K=K_{ab}\,g^{ab}=-\frac{1}{\sqrt{g}}\,\sqrt{g}^{\,\,\prime}$,
a prime meaning derivative with respect to $l$,
$K_{ab}$ being the extrinsic curvature of the two-surface,
and 
the spatial energy-momentum tensor on the layer $\Theta_{ab}$ is
$8\pi
\Theta_{ab}=K_{ab}+\left( \frac{N^{\prime }}{N}-K\right) g_{ab}$.
These quantities include matter
as well as gravitational fields.

We want to integrate the first law of thermodynamics
as given in Eq.~(\ref{1stlawdensity})
to obtain the entropy $S$ of the system
in a quasiblack hole state.
The need for
equations of
state are avoided
because
we use the quasiblack hole approach. 
Again,
we choose a sequence of configurations such that all members
remain on the threshold of horizon formation and then
integrate over this
very subset, the answer must be once again model independent.
Several conditions must be met.
The quasiblack hole limit means that
$N\to0$.
To have a regular horizon to 
an outside observer one has 
$K_{ab}=k_{ab}\,l+O(l^{\,2})$,
where $k_{ab}$ is some constant tensor. 
Then the energy density ${\varepsilon}$
remains finite, indeed
${\varepsilon}=\frac{K}{8\pi}$.
The spatial stresses, $\Theta _{ab}$ given by 
$\Theta _{ab}=\frac{1}{8\pi}
\left( K_{ab}+
\left( \frac{N^{\prime }}{N}-K\right) g_{ab}\right)$,
diverge due to
the term $\frac{N^{\prime }}{N}$, and $N=0$
in the quasiblack hole state.
In the outer region we have
seen that $N^{\prime }= \kappa$, where $\kappa$ is the
surface gravity.
So the 
dominant contribution to the first law
given in Eq.~(\ref{1stlawdensity}) comes from
the term 
$\frac{\Theta
^{ab}}{2}\sqrt{g} \,dg_{ab}$, which
then becomes 
$d(\sqrt{g}s)=\frac{\kappa }{16\pi
T_{0}}\sqrt{g}g^{ab}dg_{ab}$.
Take again into account that near the
quasihorizon the backreaction of quantum fields becomes divergent
unless one makes the choice
that $T_0$ is the Hawking temperature 
$T_{\mathrm{H}}$, i.e., 
$T_{0}=T_{\mathrm{H}}=\frac{\kappa }{2\pi}$. Thus
the first law is now
$d(s\sqrt{g})=\frac{1}{4}\,\,d\sqrt{g}$, and so the entropy density
at
the quasihorizon
of a quasiblack hole is $s\sqrt{g}=\frac{1}{4}\,\,\sqrt{g}$, up to a
constant which we put to zero.  Upon integration over the surface,
i.e., $\int d^{2}x$, we obtain
$S=\frac{1}{4}\,A_+$,
where again $A_+$ is the area of the horizon of the quasiblack hole,
for details see\cite{lemoszaslaentropyquasibhs1}.
This is the Bekenstein-Hawking entropy for a black hole.

\subsection{Entropy of electrically charged quasiblack holes}

The entropy of electrically charged spherically
symmetric quasiblack holes
can also be dealt with using the quasiblack hole approach
and the calculation follows the same lines as above,
see also\cite{lemoszaslaentropyquasibhs1}.
In particular on can calculate the entropy of
a electrically charged thin shell with a Minkowski interior
and a Reisnner-Nordstr\"om exterior. 
The electric thin shell also has, through the integrability
conditions of the first law of thermodynamics, an
exact thermodynamic solution, which yields
$S=\frac{A_+}{4}$, when the temperature of the shell
is the Hawking temperature.
When the radius of the shell is put
to its own gravitational radius, then the entropy
of this quasiblack hole is the Bekenstein-Hawking
entropy, for details see\cite{lemosquintazaslavskii1503}.

\subsection{Entropy of other quasiblack holes: rotating black holes
in three dimensions, and black holes in $d$ dimensions}

The quasiblack hole approach can be applied to a number
of other situations.
The thin shell approach, through its integrability conditions,
it is also of great interest in these cases. 
It has been applied for nonrotating and rotating shells
in three-dimensional anti-de Sitter spacetimes and,
upon taking the quasiblack hole limit,
one find the entropy of
the corresponding BTZ black hole, for details see
\cite{lemosquinta1403,lemosminami1508}.
The entropy of static shells in $d$-dimensions and the
quasiblack hole limit has also been analyzed,
see\cite{andrelemosquinta}.

\section{Entropy of extremal quasiblack holes and implications to the
entropy of extremal black holes}

The ultimate test of the quasiblack hole approach and formalism is the
study of the entropy of
quasiblack holes
in the extremal case.  This is because the entropy
of extremal black holes is a particularly intriguing and interesting
problem. Arguments based on the periodicity of the Euclidean section
of the black hole lead one to assign zero entropy in the extremal
case, $S=0$. This value $S=0$ is obtained because the Euclidean time,
and so the temperature, is not fixed in a classical calculation of the
action for extremal black holes, indeed $T$ or its inverse $\beta$ can
take any value, and this forces a zero entropy value
in the path-integral
action approach\cite{haw,teit}.  However, extremal black hole
solutions in string theory typically have the conventional value given
by the area formula $S=\frac{A_+}{4}$, a value that is obtained from
counting string states of a black hole within string theory
\cite{vafa}.  Up to now the issue has not been settled.
Neither it has
been showed that $S$ should be nonzero for extremal black holes nor it
has been shown that it should indeed be $S=\frac{A_+}{4}$.  The
situation remains to be clarified.

The quasiblack hole approach yields a method of facing this problem
and proposes a solution on the basis of pure thermodynamics. We use the
quasiblack hole approach and
we also use the thermodynamic exact solution of a thin shell to find a
thermodynamic solution for the entropy.

Let us apply the quasiblack hole approach for extremal systems.
The metric is given in Eq.~(\ref{metricgeneral1}),
and for the outside
one has $\psi=0$
and $N^2=\left(1-\frac{r_+}{r}\right)^2$, where
$r_+=m=q$, since we are dealing with the extremal case.
The electric potential is written generically as $\phi
(r)=\frac{r_{+}}{r}$ up to a constant. 
The first law given in Eq.~(\ref{1stlaw11})
in terms of boundary values
is now generalized
to include electric charge
\begin{equation}
TdS=dE+p\, dA-\varphi\, dq\,,  \label{1}
\end{equation}
where $\varphi$ is the
thermodynamic electric potential at the
boundary
and $q$ its electric charge, 
and the other quantities were defined previously.
Note that $p$ is a thermodynamic
tangential pressure defined
in\cite{by}.

We go through
the terms
$dE$, $p\, dA$, and $-\,\varphi\,dq$
carefully in order to understand the term $TdS$ in the end.
First, we deal with $dE$.
Clearly, in the extremal case $E=r_+$, so $dE=dr_+$.
Second, we deal with $p\, dA$.
One can show that 
$8\pi p R= \dfrac{4\pi
p_{r}^{\mathrm{\,matter}}R^{2}}{\left(1-\dfrac{r_{+}}{R}\right)}$.
It is clear
that $p$ is a two-dimensional pressure and
$p_{r}^{\mathrm{\,matter}}$ is a three-dimensional pressure,
and also that $p$ is a blue shifted pressure to $R$.
Now, to make
progress we have to understand the system at the threshold of being a
quasiblack hole. We have to take into account that on the quasihorizon
$p_{r}^{\mathrm{\,matter}}(r_{+})=0$ according to our general results
on pressure. When matter is absent in the
inner region, as in a thin shell, this condition is exact. When there
is matter, one can write quite generally $p_{r}^{\mathrm{\
\,matter}}(R)=\frac{b(r_{+},R)}{4\pi R^{2}}\left(
1-\frac{r_{+}}{R}\right) $, valid near $R=r_{+}$ and with the function
$b(r_{+},R)$ being
model-dependent.  The point here is that when
the body is sufficiently compressed
it follows that $p_{r}^{\mathrm{\,matter}}(r_{+})=0$. Thus, 
$
p =\frac{1}{8\,\pi }\,\frac{b(r_{+},R)}{R}\,. 
$
Now,  the area $A$ is defined as $A=4\pi R^{2}$, so that
$
dA=8\pi R\,dR$, and for $R=r_+$ we have $
dA=8\pi r_+\,dr_+$.
Third, we deal with $-\,\varphi\,dq$.
In the first law, 
the thermodynamic
electric potential $\varphi$  is
the difference between
the electric potential $\phi_{0}$ at a  reference point 
and  the electric potential $\phi (R)=\frac{q}{R}$
at the boundary $R$, blue-shifted
from infinity to $R$ through the
factor $\dfrac{1}{\left(1-\dfrac{r_{+}}{R}\right)}$,
i.e., 
$\varphi =
\dfrac{\left(\phi_{0}-
\frac{r_+}{R}\right)}{\left(1-\dfrac{r_{+}}{R}\right)}
$.
Now for $R$ near $r_+$
one has $\phi_{0}\to 1$ so we can put
$\varphi =f(r_+,R)$.
For $dq$ in the first law,
since at the quasihorizon limit $q=r_{+}$, one has
$
dq=dr_{+}  
$.
We are now ready to analyze the entropy of
extremal quasiblack holes.

We assume that the integrability conditions
for the system are valid, otherwise there is no thermodynamic
system. Then, since $S$ is a total differential one can integrate
along any path. Choose the path $R = r_{+} (1 + \delta)$ with $\delta$
constant and small, so that from what we found above we get
the first law~(\ref{1}) in the form 
$TdS=\left(1+b_{+}-f_{+}\right)
\,dr_{+}$. Clearly, the
integrability condition yields that
the local temperature $T$ is of the form
$T=T(r_+)$.
Then, 
$dS=\frac{\left(1+b_{+}-f_{+}\right)}{T(r_+)}
\,dr_{+}$
where,
$b_{+}=b(r_{+}, R=r_{+})$ and $f_{+}=f
(R=r_{+})$.
One can now integrate this equation to obtain $S$, already
with the limit taken
$R\rightarrow r_{+}$, to obtain
\begin{equation}
S=S(r_{+})=\int_{0}^{r_{+}} d\bar{r}_{+}\,\frac{D(\bar
r_{+})}{T(\bar{r}
_{+}) } \,,  \label{entropy}
\end{equation}
where, $D(r_{+})= 1+b_{+}-f_{+}$.  In general,
one should require only
that
$1+b_{+}-f_{+}>0$ to ensure the positivity of the entropy. Note that
if the density of matter inside vanishes for $r<R=r_+$, we have a thin
shell situation, $b_{+}\rightarrow 0$ and $f_{+}\rightarrow 0$, and
so $D(r_{+})=1$.

Now, the local temperature $T$ at the boundary $R$ is in general a
function of $r_+$ and $R$, $T=T(r_+,R)$, and it is related to the
temperature $T_{0}$ at infinity by the Tolman formula,
$T=\frac{T_{0}}{N(r_+,R)}$, i.e., $T=\frac{T_{0}}{1-\frac{r_{+}}{R}}$.
But we
have just
deduced above that in the extremal case at the quasiblack hole limit
$T$ is a function of $r_+$ solely, and not of $R$, $T=T(r_+)$.  So,
from the Tolman formula, the temperature at infinity $T_0$ has thus
the form,
\begin{equation}
T_0=T(r_{+})\left(1-\frac{r_{+}}{R}\right)\,,
\label{temptolman1}
\end{equation}
and therefore 
$T_0=T_0(r_+,R)$.
With Eqs.~(\ref{entropy}) and~(\ref{temptolman1}) in hand we can now
draw several conclusions relatively to the entropy of extremal
quasiblack holes.

One possible case,
from Eq.~(\ref{entropy}), is the one
that has finite generic local temperature $T(r_{+})$, the entropy
$S(r_{+})$ of quasiblack holes is positive, $S(r_{+})>0$, and can be
any well behaved function of $r_+$. Moreover, in this case, the
temperature at infinity $T_{0}$, see Eq.~(\ref{temptolman1}), goes to
zero in this limit, $T_{0}\to0$.

Another possible case in this approach is when the local temperature
$T$ behaves as $T(r_{+})\to\infty$ when one assumes $T_0$ finite,
i.e., $T_0\neq 0$, see Eq.~(\ref{temptolman1}).  This means that the
local temperature $T(r_{+})$ is infinite for every $r_+$ in the
integration process of Eq.~(\ref{entropy}) and so one obtains $S=0$.
This case of extremal quasiblack hole behavior is equivalent to the
path integral prescription given in\cite{haw,teit} for extremal pure
black holes.
It is known that the quantum
stress-energy tensor of fields in a black
hole spacetime 
blows up when the local temperature $T$ goes to infinity,
and so this case, if accepted, somehow
avoids this problem.

Thus, on taking into account the two cases, i.e., $T_0=0$, and
$T_0>0$, altogether, one can say
that the entropy $S$ is a function of
$r_+$, although an undetermined function, $S=S(r_+)$.  In the extremal
case the stresses are finite, and so one can deduce that not all
possible modes are excited when the quasiblack hole state is
approached.  Since for nonextremal quasiblack holes $S=\frac14 A_+$
and all the possible modes due to the
infinite stresses are excited here, one
concludes that the entropy of extremal quasiblack holes should be
$S\leq\frac14 A_+$. Changing the variable $r_+$ to $A_+$, we obtain
that the entropy of an extremal quasiblack hole is $S=S(A_+)$ with
$S(A_+)$ arbitrary, bounded from below by $0$ and from above by
$\frac14 A_+$, i.e.,
\begin{equation}
0\leq S(A_+)\leq\frac14 A_+\,.
\label{s2}
\end{equation}
In brief, we showed consistently that the thermodynamic treatment does
not give an unambiguous universal result for $S(A_+)$. The entropy
depends on the properties of the working material and, moreover, on
the manner the temperature approaches the zero value. In particular,
$S=\frac{A_+}{4}$ is not singled out beforehand for the extremal
black hole entropy.  This holds for extremal quasiblack holes and by
inference it holds for extremal black holes.  So, our approach points to the
conclusion that the extremal entropy depends on the manner the
extremal quasiblack hole, and thus the extremal
black hole, has formed, for details
see\cite{lemoszaslavskiiextremalqbhentropy}, see also
\cite{pretvollist}.

There is another important approach for the thermodynamics and entropy
of extremal quasiblack holes that yields exact results.  It is the
thin shell approach, which is an exact solution as a general
relativistic system and has an exact solution as a thermodynamic
system. 
For an extremal shell one finds that there are three distinct
cases.  First, one starts with a nonextremal shell, studies its
thermodynamics, puts it at the gravitational radius, and turns it
extremal there.  Not surprisingly, the entropy is $S=\frac14 A_+$, in
this case only the pressure term contributes and all the modes have
been excited. Second, one turns the shell extremal
concomitantly with its approaching of its own gravitational radius.
Surprisingly, one
obtains $S=\frac14 A_+$, and in this case all forms of energy in the
first law of thermodynamics contribute to give this value. Third, one
turns the shell extremal and only afterwards one approaches the
gravitational radius. One finds here that $S=S(A_+)$, the entropy is
any well behaved function of $A_+$.  So can conclude again, now from
the thin shell solution approach
and its three distinct cases, that $0\leq S\leq\frac14 A_+$,
i.e., Eq.~(\ref{s2}) holds. This result shows
that the quasiblack hole and the
thin shell approaches are consistent.
For
details see
\cite{lemosquinzaslaextremalentropy1,lemosquinzaslaextremalentropy2}.
For extremal three-dimensional rotating black holes in anti-de Sitter
spacetimes see\cite{lemosminami1701,lemosminami1709}, where the
same type of results and conclusions are drawn.

\section{Conclusions}

Quasiblack hole solutions are matter solutions up to a boundary $R$
which is a quasihorizon, i.e., $R=r_+$.  A quasiblack hole is a
regular solution in the sense that the Kretschmann scalar is finite
everywhere, although there is some form of degeneracy at the horizon,
namely, for external observers the horizon is a naked singular
horizon. This degeneracy, that combines features typical of regular
and singular systems, is made clear in the Carter-Penrose diagram of a
quasiblack hole, where the interior and exterior regions are disjoint.
The pressure properties of a quasiblack hole at the quasihorizon can
be calculated and regularity imposes very strict conditions on them. A
mass formula for quasiblack holes can be obtained, which is identical
to the mass formula for black holes, although derived from totally
different techniques. In studying the entropy of a nonextremal
quasiblack hole we recover the Bekenstein-Hawking entropy.  The result
and the way it is established suggest that the degrees of freedom are
on the horizon, since it is when a horizon is formed and the system
has to settle to the Hawking temperature that the entropy takes the
value $S=\frac{A_+}{4}$.  The results also suggests that the degrees
of freedom are gravitational modes, since when the nonextremal
quasiblack hole state is approached the tangential pressure goes to
infinite or, more appropriately, to the Planck pressure.  Then modes,
presumably quantum gravitational modes, are induced.  The difficult
issue of the entropy of extremal black holes is a high point of the
quasiblack hole approach. One finds that the entropy is a generic
function of $A_+$, $S=S(A_+)$, and the precise function depends on the
manner the quasiblack hole has been formed. Moreover, the entropy of
extremal quasiblack holes, and thus extremal black holes, should be
bounded by the Bekenstein-Hawking value $\frac{A_+}{4}$, so that
$0\leq S\leq\frac{A_+}{4}$.

Several final remarks can be drawn.
First, quasiblack holes are stars at the quasiblack hole limit and as
such can be considered the genuine frozen stars,
now that the frozen star name endorsed in the past for black holes
has not this specific use anymore.
Second, quasiblack holes, not black holes, are the real descendants of
Mitchell and Laplace stars.
Third, quasiblack holes are Schwarzchild and Buchdahl stars pushed, by
use of some added repulsive charge, to their maximum compactification,
i.e., the trapped surface limit.
Fourth, in this sense, quasiblack holes are objects on the verge of
becoming black holes, and as such can be envisaged as a metastate of
spacetime and matter. Continued gravitational collapse ends in black
holes, whereas quasistatic contraction passes through a quasiblack
hole phase. Fifth, quasiblack holes have
special properties, and their Carter-Penrose diagrams
manifestly incorporate features of normal
stars, regular
black holes, and null naked
horizons.
Sixth, the quasiblack hole approach, in some of its aspects, is a
cousin of the membrane paradigm.  By taking a timelike matter surface
into a null horizon we are retrieving the membrane paradigm. One
difference is that the quasiblack hole membrane is not fictitious like
in the membrane paradigm, it is made of real matter.
Finally, studies to understand the entropy of systems with other types
of horizon can be taken with the quasiblack hole approach.

\section*{Acknowledgments}
We are thankful to our collaborators Rui Andr\'e, Kirill Bronnikov,
Francisco Lopes, Masato Minamitsuji, Gon\c calo Quinta, Jorge Rocha,
Erick Weinberg, and Vilson Zanchin.  We are thankful to Luis
Crispino for, on inviting us to present this work in the V Amazonian
Symposium on Physics in Bel\'em do Par\'a in November 2019, speeded
us to write this review and enabled us to present
the Carter-Penrose diagrams for quasiblack holes, two
assignments that were on our
list for a long time. JPSL acknowledges Funda\c c\~ao para a Ci\^encia
e Tecnologia FCT - Portugal for financial support through
Project~No.~UIDB/00099/2020. OBZ thanks Kazan Federal University for
a state grant for scientific activities.

\vfill


\end{document}